# Back-end Electronics for Low Background and Medium Scale Physics Experiments Based on an Asymmetric Network

D. Calvet

*Abstract*—The detector readout architecture introduced in this paper is intended for small to medium size physics experiments that have moderate bandwidth needs, and applications that require an ultimately low background radioactivity for the parts close to the detector. The first idea to simplify the readout system and minimize material budget is to use a common fanout structure to transport from the off-detector back-end electronics all the traffic required for the synchronization, configuration and readout of the front-end electronics. The second idea is to use between each front-end card and the back-end electronics a point-to-point link that runs at the relatively low speed that suffices for the target application. This broadens the possible choices for the physical media of the communication links, e.g. glass fiber, plastic optical fiber, or copper.

This paper presents a communication protocol adequate for the proposed asymmetric network and shows the design of a back-end unit capable of controlling 32 front-end units at up to 12.8 Gbps of aggregate bandwidth using an inexpensive commercial FPGA module where the large number of regular I/O pins interface to the front-end links, while the few available multi-gigabit per second capable transceivers are affected to the communication with the upper stage of the DAQ system.

## I. INTRODUCTION

SOME of the recurrent questions that arise when designing the data acquisition system of a physics experiment, are how to distribute a common clock and synchronization signals to multiple front-end boards, how to download run time parameters in each of them, how to collect experimental data and how to transport slow control and monitoring traffic. Many solutions have been proposed: distinct or common networks for synchronous signals and asynchronous data, standard or custom protocols, etc.

The availability of high speed SERDES in modern FPGAs has considerably simplified the way to implement communications links in a data acquisition system. Although large physics experiment have a constant thirst for higher bandwidth, many small and medium scale experiments do not need ultimate transfer speeds. For example, low radioactivity background experiments, like the PandaX-III neutrinoless double beta decay search experiment [1] is expected to produce less than 1 Gbit/s of aggregate data. This translates into a modest required bandwidth of 50 Mbps for each of the 21 front-end cards used to read out the 5000 channels of each Time Projection Chamber (TPC) end-plate. However, this application has stringent requirements in terms of low radioactivity for the electronics close to the sensitive detection volume. Minimizing material budget and carefully selecting components for their low contamination of radioelements is of primary importance. Another application, the readout of the TPCs in the upgraded T2K neutrino oscillation experiment [2], will comprise 30,000 electronic channels, but the total amount of data to record is expected to be less than 10 MByte/s.

The goal of this paper is to build a flexible detector readout architecture where communication links are tailored to the real bandwidth needs of medium-scale applications like the above ones. This increases the range of possible optimization for low background radioactivity on the front-end side (if this is needed), and it also simplifies the design of back-end electronics which can be implemented around a commercial, cost-effective, System-on-Chip (SoC) FPGA module.

## II. DESCRIPTION OF THE CONCEPT

A schematic view of the proposed detector readout system is shown in Fig. 1. For minimal signal degradation, front-end cards are placed close to detectors. Reduced material budget, low power consumption, limited space and accessibility are typical constraints for these. Radiation hardness is not needed for the two applications previously mentioned. Back-end electronic boards are placed several 10 meters away from the front-end and do not have constraints in terms of low background radioactivity, radiation hardness, or limited access.

The information to transport from the back-end side to the front-ends consists of : 1) a global synchronization clock, 2) a global trigger signal and some other synchronization information (e.g. event type, event number, timestamp), 3) configuration data for the front-ends, 4) periodic slow control monitoring request, 5) messages to throttle the transmission of data acquired from the front-ends. The trigger rate for the target applications will not exceed ~10 Hz. This translates into an extremely small bandwidth needed to transfer trigger information, e.g. 1 kbps assuming that each trigger message is 100 bits. Network transit latency is not critical for externally triggered slow detectors like TPCs. Because the primary function of the back-end to front-end path is to distribute a common clock and synchronous signals, a fanout topology is the most natural solution. In order to limit material budget and make efficient resource usage, the idea is to also use this fanout link to transport traffic 3), 4) and 5). Configuration data need only be downloaded once in the front-end at startup (few kB). Monitoring requests are typically sent every few seconds. Messages used for controlling data acquisition do not require very high bandwidth either. The speed of the fanout is therefore mostly determined by the frequency of the global clock to distribute (a submultiple could be send) and the desired time resolution for synchronization. In the reported system, the reference clock is 100 MHz, and a speed of 100 Mbps for the back-end to front-end fanout tree was chosen.

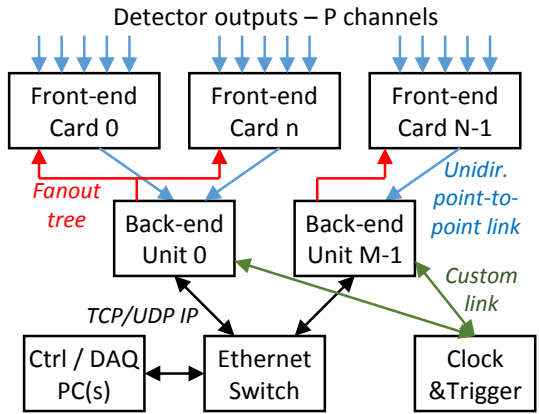

Fig. 1. Block diagram of the proposed detector readout architecture. The communication from each back-end unit to its set of front-end cards relies on a fanout tree while point-to-point links run in the opposite direction. Standard Ethernet networking is used between back-end units and the DAQ/control PC(s). A set of custom bi-directional point-to-point links is used for clock and trigger distribution to the back-end units.

In the front-end to back-end direction, a point-to-point unidirectional link per front-end card is used. Assuming that each of the N front-end cards produces the same amount of data on average, the mean required bandwidth for a front-end link is simply 1:Nth of the global DAQ bandwidth. A typical medium-size application that records ~20 MB/s of data gathered from ~20 front-end cards, only needs to have 8 Mbps capable front-end links. Low speed links broadens implementation options which helps when optimizing for some other constraints, e.g. low background radioactivity. In addition, regular FPGA I/O pins can be used on the back-end side to de-serialize data. This provides a more compact and cost effective implementation than embedded high speed SERDES. Low speed links induce longer data transfer times, but if the front-ends can buffer multiple events, this does not increase system dead-time.

The present system is sized for 1 Gbps of maximum global DAQ sustained throughput, i.e. 10 TB of data per day. This is amply sufficient for many small and medium scale experiments. The nominal speed selected for the front-end to back-end links is 400 Mbps to cover a wide range of possible applications.

## III. COMMUNICATION PROTOCOL

Time-division multiplexing with a fixed, pre-defined cyclic pattern is used to transport simultaneously synchronous and different types of asynchronous traffic between the front-end and back-end sides. The physical link bandwidth is divided into three virtual channels of unequal bandwidth as shown in Fig. 2.

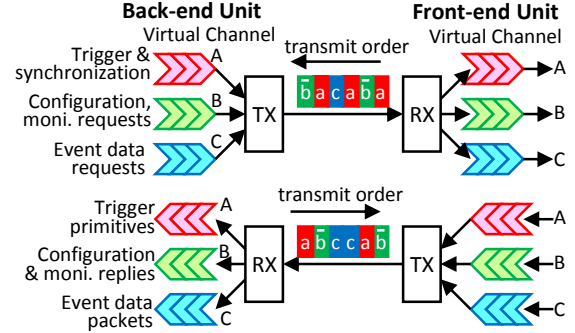

Fig. 2. Time-division multiplexing for the simultaneous transmission of 3 different classes of messages. In the back-end to front-end direction, one bit from each virtual channel is sent in turn following the fixed sequence: A, B, A, C. In the opposite direction, the transmission sequence takes one bit successively from: A, B, C, C.

In the back-end to front-end direction, virtual channel A is used to transport the synchronous trigger signal and related information. In the opposite direction, virtual channel A carries trigger primitives (if this function is used), a trigger acknowledge flag and a busy flag. Virtual channel B is used for front-end run-time parameters download, read-back, and for slow control monitoring. Virtual channel C is used for the acquisition of detector data and the associated protocol. In the back-end to front-end direction, the bandwidth is split in 50%, 25%, and 25% for virtual channel A, B and C respectively. In the opposite direction, the share is 25%, 25% and 50%. Half of the link bandwidth (i.e. 200 Mbps) is reserved for the transport of experimental data. Message delivery with deterministic latency can easily be achieved using static time division multiplexing. This scheme is simpler to implement than other methods like creating several levels of priorities for different classes of messages or designing a mechanism to interrupt, and later resume, the transmission of an asynchronous message to insert a message with deterministic latency constraints.

### A. Message Format

The format of messages sent over virtual channel A in the back-end to front-end direction is shown in Fig. 3. In the opposite direction, a comparable format is used. The message contains a SET_BUSY bit used to acknowledge the trigger, a CLEAR_BUSY bit used to indicate read out completion, and four optional trigger primitive bits that are used when the system operates in self-trigger mode.

Each front-end card has some register space for storing runtime parameters and monitoring data. Logically, it is organized as 64 KB of memory space that is made accessible to the back-end electronics over a 16-bit address / 32-bit data virtual bus. Each read or write request sent from the back-end unit is serialized over virtual channel B and must be echoed by a response message from the front-end card target(s).

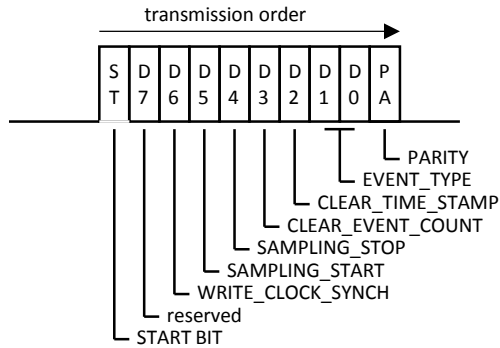

Fig. 3. Format of message from back-end to front-end over virtual channel A. The 8-bit payload comes after a start bit and is followed by a parity bit. The trigger is the SAMPLING_STOP bit. The EVENT_TYPE bits allow distinguishing four types of triggers. Data sampling is resumed using the SAMPLING_START bit. Other bits are used to clear the local event counter, event timestamp counter, and synchronize the local generator that produces the clock used for sampling detector signals.

The format of messages on virtual channel B is shown in Fig. 4. Series of read and write operations are made over the virtual bus to download run time parameters in each front-end card and retrieve slow control monitored variables during operation.

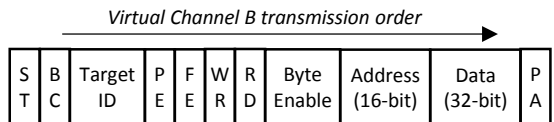

Fig. 4. Format of messages over virtual channel B. Each message is composed of a start bit (ST), 62 payload bits and one parity bit (PA). The 5-bit Target ID field determines on which front-end card, numbered from #0 to #31, the transaction takes places. A broadcast bit (BC) indicates that the transaction is for all front-end cards. The type of transaction, read or write, is determined by the RD and WR bits respectively. Four byte enable bits specify which bytes on the 32-bit data lines are active. The 16-bit address of the transaction is supplied along with 32 bits of data. In the front-end to back-end direction, the PE bit is used to indicate a parity error on the received request and the FE bit is used to signal a local bus error. For read operations, the data field contains the read data. For write transactions, it echoes the write data of the request.

The purpose of virtual channel C is to transport the data acquired by the front-ends to the back-end side. The format of a request message and response packet are shown in Fig. 5(a) and Fig. 5(b) respectively. Data transfers are initiated by requests sent from the back-end to the front-ends. A fanout structure is inefficient for distributing different individual messages to each front-end because these have to be sent sequentially. To overcome this limitation, messages sent by the back-end over virtual channel C specify multiple targets for the desired operation using unary coding: in addition to the code of the operation itself, the message contains a 32-bit field where each bit set in this field indicates that the front-end card with a matching ID must execute the current operation. Hence, only one request message is needed to trigger an action in multiple and possibly all front-end cards. So far the only action that is defined instructs the target front-ends to send the next packet of event data. Other actions could be defined, for example to re-send a packet in case of a transmission error.

In the front-end to back-end direction, event data are packetized in 16-bit words that are serially transmitted (MSB first) over virtual channel C. Currently, each packet corresponds to the data of one detector channel. A packet starts with a header that contains: 1) a start of event flag (SOE), 2) an end-of-event flag (EOE), 3) the size of the packet in bytes. The packet header is followed by an even number of data payload words and cyclic redundant code (CRC-32) for error detection. In the current implementation, corrupted packets are not retransmitted. An event is flagged as incomplete when one or several fragments have been dropped because of a CRC error.

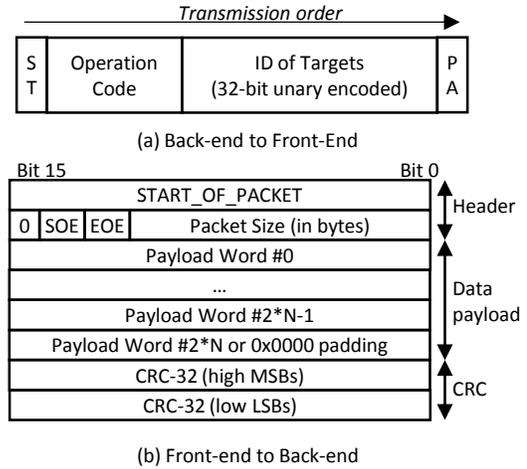

Fig. 5. Virtual channel C data request message from back-end to front-end (a) and response event fragment packet (b).

### B. Front-end Card ID Assignment Procedure

The ID of every front-end card is assigned at system startup by the back-end board to correspond to the physical port, numbered from #0 to #31, where each front-end card is connected. Because a fanout network is used in the back-end to front-end direction, the back-end board cannot assign the ID of each front-end card directly at cold start because addressing front-end cards individually precisely relies on each card ID being set. The following procedure is used to bootstrap communication. In a first step, the back-end board reads a unique serial number hardcoded in each front-end card. This is accomplished over virtual channel B with the broadcast target bit set. In a second step, the back-end board sends in broadcast mode the list that gives the correspondence between the unique serial number of each front-end card and its assigned port ID. Each front-end card detects its own unique serial number in this flow and captures its assigned ID. In this design, card serial numbers are obtained with the 53-bit DNA number hard-coded in the silicon of the Xilinx Artix 7 FPGA of each front-end card.

### C. Line Encoding for the Back-end to Front-end Fanout

Using a fanout structure for the back-end to the front-end communication brings the drawback that when the back-end transmitter is reset, all front-end receivers are affected. It is therefore highly desirable that receivers are capable of gaining synchronization on-the-fly without the need to interrupt the transmitter. Each receiver must also be capable of delineating the bits of the different virtual channels. Finally, a DC balanced protocol should be preferred to keep the largest possible choice of physical media. The encoding method shown in Fig. 6 meets all these requirements.

Firstly, the data of virtual channel B (it could be C) are inverted. In the absence of any message being sent (marked by sending 0's), the receiver can easily delineate the serial stream of each virtual channel: the bit equal to 1 corresponds to the inverted virtual channel B, the previous and following bits map to virtual channel A, and the next following bit originates from virtual channel C. The second step is Manchester encoding: every original bit is converted into two bits with half duration: the original bit followed by its inverted value. Although the baud rate on the media becomes twice the actual link bit rate, speed remains low (i.e. 200 Mbaud for the original 100 Mbps bit stream).

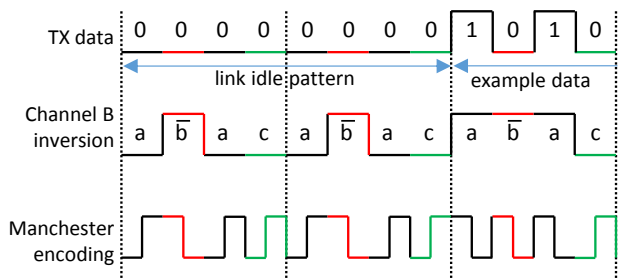

Fig. 6. Serial line encoding for the back-end to front-end fanout link. Firstly, the bits to transmit over virtual channel A, B and C are interleaved following the sequence {a, b, a, c}. Secondly, the bits from virtual channel B are inverted. Thirdly, each bit followed by its inverted value is transmitted over the physical media. When all virtual channels are idle, this leads to the constant pattern "01100101" being sent. This pattern is used to maintain link synchronization, to determine the time for optimal bit capture and to perform the delineation of the 3 virtual channels.

When no traffic flows through the media, a fixed constant pattern "01100101" is permanently sent by the back-end transmitter. Receivers are required to de-serialize this bit stream and perform bit slip operations until synchronization to this pattern is gained.

### D. Line Encoding for the Front-end to Back-end Links

In this direction, each transmitter-receiver couple can be reset independently without affecting other links. A simple encoder and decoder (DC-balanced) is needed as well as a simple method for virtual channel delineation. It is intended to de-serialize data using regular FPGA I/O pins on the back-end side. Because such pins do not have clock recovery capability, the local global clock is used. This type of receiver is referred to as "a system synchronous de-serializer". For correct operation, it is mandatory that every front-end transmitter is synchronous to the global system-wide clock that is recovered by its receiver. After being reset, a front-end transmitter is required to send a pattern of alternating ones and zeroes at 400 Mbps during 100 ms. This training period allows the back-end receiver to adjust the delay applied to the received stream to sample it at the optimal time with the global clock. This delay calibration is currently performed only once after the link is reset, but dynamic tracking of skew could be implemented. Note that system synchronous receivers are sensitive to the sum of the latency variations of the forward and return path of the link. After the initial training period, the front-end transmitter switches to the transmission of the three time interleaved virtual channels. The data of virtual channel B are inverted before transmission for easier delineation on the receiver side. Because event data are acquired in this link direction, a method that has low encoding overhead is preferable. Using 8B/10B encoding would be possible, but regular FPGA I/O pins do not support natively common standard protocols. Implementing an array of 8B/10B decoders in the FPGA fabric would consume some resources. For simplicity and efficiency, a self-synchronizing scrambler is used. It is based on the polynomial $x^{43}+1$ from the ATM standard [3]. As shown on Fig. 7, both the encoder and the decoder are simple. No bandwidth is wasted by encoding.

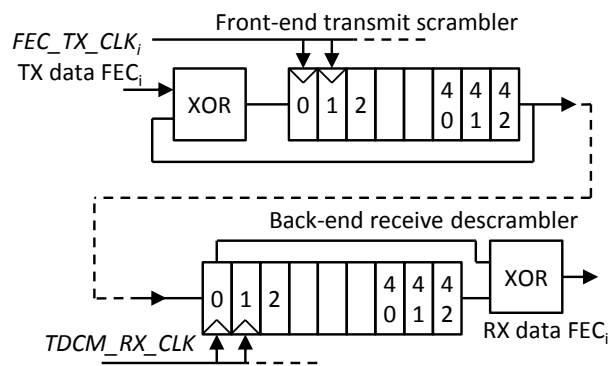

Fig. 7. Self-synchronizing scrambler and descrambler used for the front-end to back-end links. The front-end transmitter must be synchronous to the recovered clock of the receiver which is also synchronous to the system-wide common clock.

## IV. IMPLEMENTATION

### A. Front-end Card

The ARC is a general purpose 6U form factor front-end card compatible with the proposed back-end system. It was built in the framework of the Harpo project [4]. The ARC supports the AFTER chip [5], the AGET chip [6] and the ASTRE chip [7]. The ARC provides 256 channels of charge sense amplifiers (120 fC to 10 pC range), shapers with programmable peaking time (100 ns to 8 µs) and 512 time-bucket switched capacitor array. The ARC contains the required ADC, FPGA, event buffer SRAM, and ancillary circuitry for configuration and readout over a serial optical link using the protocol previously described. Standalone operation with readout by a PC via Gigabit Ethernet is also supported.

Various techniques have been proposed to achieve deterministic latency with high speed transceivers embedded in FPGAs [8], [9], [10]. However, operation at low speed, under 500 Mbps, is outside of the acceptable range of the GTP transceiver of the Xilinx Artix 7 FPGA used on the ARC. To work around this limitation, a clock and data recovery (CDR) chip external to the FPGA is used. Two commercial devices have been investigated: the ADN2815 from Analog Devices (10 Mbps to 1.25 Gbps) and the SY87701AL from Micrel (28 Mbps to 1.3 Gbps). The datasheets of these two commercial devices do not provide a characterization of the input to output delay. Test boards have been built to perform this measurement. No measurable difference of delay between the recovered clocks of two SY87701AL chips was observed, but up to 1.8 ns skew was measured between two ADN2815. However, for both parts, the skew remains constant from one power cycle to the next and does not appear to change over time. On the ADN2815, a drift of 620 ps for the phase of the recovered clock compared to a reference device for a variation of the ambient temperature from 0°C to 50°C was found. No variation was seen over a power supply change from 3 V to 3.6 V. The ADN2815 was retained for the ARC because the recovered clock and serial data pins use LVDS. This interfaces directly to I/O pins of the local FPGA. A similar design based on the SY87701AL would require adding PECL to LVDS translators.

To achieve deterministic latency, the CDR must feature a constant delay, but this is not sufficient. Using Manchester encoding, a 100 Mbps bit stream leads 200 Mbaud on the physical media. The frequency of the clock recovered by the CDR chip is therefore 200 MHz. The receiving FPGA performs a division by 2 to retrieve the original 100 MHz clock, but this process leads to two possible 100 MHz clocks, 180° apart. To resolve the corresponding 5 ns uncertainty, the receiving FPGA shall de-serialize data with the 100 MHz clock that produces the nominal idle link pattern "0100" rather than the out-of-phase clock that produces the inverted idle link pattern "1011". A receiver composed of a CDR (internal or external) and the associated FPGA logic is referred to in this paper as "a source synchronous de-serializer" as opposed to the "system synchronous de-serializer" defined earlier in the text.

*B. Back-end Unit*

The back-end board that implements the concepts described in this paper is called the "Trigger and Data Concentrator Module", TDCM. It is composed of three part:
- A commercial SoC FPGA module that contains all the intelligence of the TDCM,
- A custom made carrier board that supports the FPGA module, provides power and all the interface connectors to the outside world,
- One or two physical layer mezzanine cards.

The selected FPGA module is the Mercury ZX1 SoC [11] from Enclustra. It contains a Xilinx Zynq 7z030/35/45 FPGA, 1 GB of DDR3 SDRAM, four or eight 6.6-10 Gbps capable serial transceivers, ~140 user I/O pins, an embedded Ethernet PHY, a USB controller, flash memory, and more.

The carrier board of the TDCM is a 6U form factor PCB. It takes a 12 V power input and houses the required DC/DC converters to power all the components of the TDCM. Four high speed GTP transceivers are routed to SFP cages, while the remaining four GTPs (only available with the 7z035 and 7z045 Zynq FPGA) are assigned to a PCI Express Gen2x4 cable interface. An RJ45 jack connects to the Gigabit Ethernet PHY and controller embedded in the Zynq. Because this controller does not support Jumbo frames, a softcore Ethernet MAC (Xilinx TEMAC) is synthetized in the FPGA fabric and is connected to one GTP. An optical transceiver for Ethernet 1000-base X or a Gigabit media converter for RJ45 copper media is inserted in the corresponding SFP cage. The TDCM has a USB to RS-232 bridge for console printout, 3 front-panel LEDs, several push buttons, a micro SD card for storing the embedded firmware and software, 6 NIM level and 3 TTL level I/O's, and two RJ45 connectors for cascading TDCMs or interfacing to some master clock and trigger device. A low pin count FMC connector is placed on each side of the TDCM carrier board to connect a physical layer mezzanine card.

The physical layer mezzanine card contains two 8-slot SFP cages. It can accommodate up to 16 bi-directional optical links. The transmit path takes 16 identical copies of the serial stream sent by the TDCM to the front-end side. The duplication is physically made by two fanout chips (Texas Instruments CDCLVD1216) placed on the TDCM carrier board. The transmit path consumes only one pair of LVDS I/O pins on the FPGA module. The receive path takes 32 LVDS I/O pin pairs. Other types of physical layer mezzanine cards (e.g. plastic optical fiber, or copper) will be designed if needed.

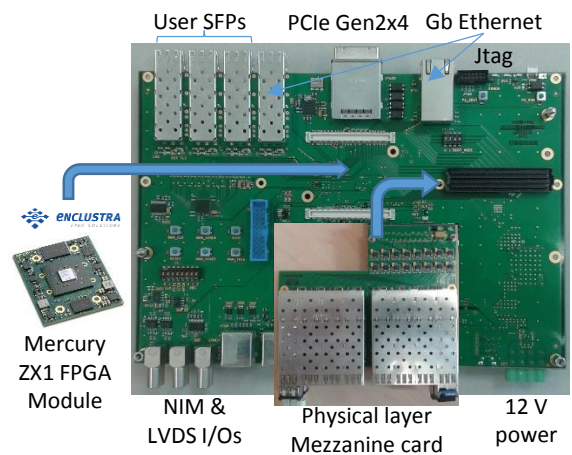

Fig. 8. Picture of the TDCM. It is composed of three elements: a custom-made 6U form factor carrier board, a commercial FPGA module, and up to two physical layer mezzanine cards. Model shown: 16 optical ports (SFP).

A picture of a TDCM is shown on Fig. 8. The width of a TDCM equipped with two physical layer mezzanine cards, is 80 mm. A standard 19" crate can house up to 5 TDCMs. The TDCM can interface to up to 32 front-end units. The total available bandwidth in the front-end to back-end direction is 12.8 Gbps (half of this bandwidth is available for event data). If more than 1 Gbps of DAQ bandwidth is needed, one or several of the 6.6-10 Gbps capable GTP transceivers could be used. The PCIe interface port (Gen2x4, i.e. 20 Gbps) is not yet tested.

*C. Embedded Firmware and Software*

In the standalone version of the ARC, the firmware of the on-board Artix 7 FPGA contains an embedded MicroBlaze softcore processor to interpret the commands received from the control PC over UDP/IP, drive the hardware as required, and send back the acquired data. In the version where control is done by the TDCM, the softcore processor is removed to minimize the footprint of the firmware and the proprietary serial protocol described earlier is implemented in the FPGA.

On the TDCM side, the required functions are partitioned between firmware implemented in programmable logic and software running on the embedded ARM processor of the Zynq SoC. The interface to serial links, handling traffic that requires deterministic latency (virtual channel A), data collection from the front-ends (virtual channel C) and packetization in Ethernet frames are mostly implemented in FPGA logic while interpreting the commands received from the control PC, performing configuration and monitoring operations (virtual channel B), pre-pending UDP/IP headers to the buffers that have been filled by the underlying hardware with acquired data, are tasks handled in software. At present, the TDCM simply runs a standalone bare-metal command interpreter program but running locally the MIDAS DAQ middleware [12] on-top of an embedded version of Linux is an evolution under consideration.

### D. Principle of Event Data Collection and Transfer

Fig. 9 shows how data collection and transfer are handled within the TDCM. The logic attached to virtual channel C of each front-end link comprises two blocks: a finite state machine, the "*DataPump*", and a first-word fall-through FIFO, "FE-FIFO", capable of storing the largest packet expected from a front-end card (2 KB in this design). The role of each *DataPump* is to post a request to its front-end partner whenever its FE-FIFO has enough room to store the next packet of data. When its FE-FIFO is filled, the *DataPump* holds its data request token until enough room is available.

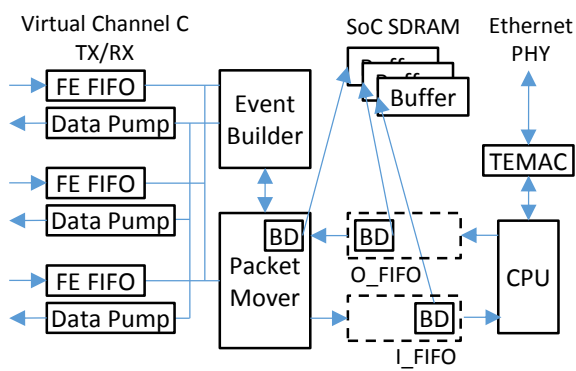

Fig. 9. Schematic diagram of front-end data reception, formatting and transfer. Data fragment collection and transfer to external SDRAM buffers are handled in firmware. The local CPU handles IP/UDP header encoding and decoding and the exchange of buffer descriptors (BD) between the firmware logic and the Ethernet media access controller (TEMAC).

The *EventBuilder* is a finite state machine that scans all FE-FIFO in round-robin until packets are available. Normally, the first packet received from each front-end card has the Start-Of-Event flag set. If not, an error is raised. Upon reception of the first Start-Of-Event packet, the *EventBuilder* copies locally the 32-bit event number and 48-bit timestamp contained in this packet. When unloading the Start-Of-Event packet of other front-ends, the received values are checked against the expected ones and the *EventBuilder* is halted in case of mismatch. After the correct Start-Of-Event packet has been received from all active front-ends, the *EventBuilder* fetches a free buffer descriptor from the O_FIFO which contains a list of buffer descriptors that point to free buffers (8 KB, i.e. one Ethernet Jumbo frame) located in the SDRAM of the SoC. A pool of free buffers is pre-allocated at system initialization and the O_FIFO is filled by the processor with the corresponding descriptors. After a free buffer is obtained, the *PacketMover* writes the header of the event in this buffer. Next, the *EventBuilder* normally receives from the front-ends the data packets that make the core of the acquired event. For each packet, the CRC-32 is verified on the fly: if it is correct, the packet is kept, otherwise it is deleted. At present, data retransmission in case of error is not done. Each time a new valid data packet is received, the *PacketMover* transfers it to the appropriate SDRAM locations using burst transfers (AXI-4 protocol). Because the size of a packet is placed in the first word of its header, the *PacketMover* can determine if the entire packet can fit in the current buffer *before* unloading any word from the FE_FIFO. If the current buffer does not have enough space, the *PacketMover* places the corresponding buffer descriptor in the I_FIFO and fetches a free buffer descriptor from the O_FIFO. Whenever the I_FIFO is not empty, the local processor unload the buffer descriptors that point to filled buffers, writes the Ethernet, IP and UDP headers at the beginning of each buffer, and passes the information to the Ethernet TEMAC hardware that performs the actual transfer to the DAQ PC. All transfers are done from the SDRAM of the SoC in DMA mode without any data copy. After a buffer has been sent, its associated buffer descriptor is returned by the processor to the O_FIFO for later use by the *PacketMover*. This process repeats until each front-end sends its End-Of-Event packet. Then, the *EventBuilder* emits a global End-Of-Event packet and starts collecting the Start-Of-Event packets of the next event.

## V. PERFORMANCE MEASUREMENTS

### A. Communication Links Between Front-ends and TDCM

The evaluation of the performance and stability of serial communication at medium speed over an optical media interfaced to FPGA regular I/O pins is the most critical aspect to validate the proposed concept.

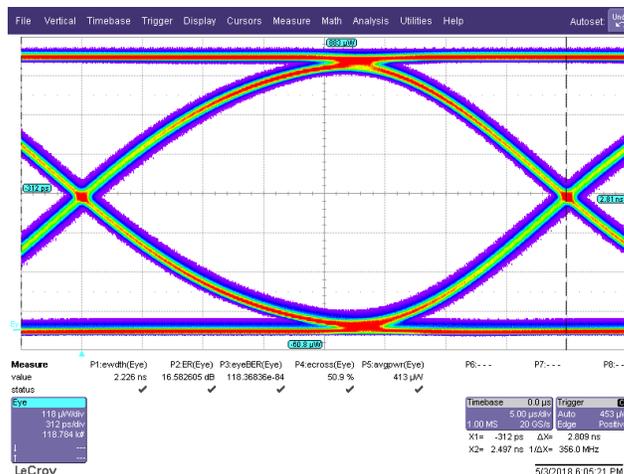

Fig. 10. Eye diagram of one of the 32 transmitter ports of the TDCM (user I/O TX with external LVDS fanout). The transmitted pattern is PRBS7 at 400 Mbps. The bandwidth of the optical converter probe of the oscilloscope is automatically limited to 350 MHz.

On the TDCM, the transmitted stream originates from a FPGA user I/O pair. It passes through the connector of the SoC module, two stages of LVDS fanout, one FMC connector and one LVDS to PECL translator before it reaches an SFP transceiver. Despite this long path, the quality of the optical signal remains very good as it can be seen on Fig. 10. This measurement was done at 400 Mbps which is twice the nominal speed used for our applications (200 Mbaud).

Fig. 11 shows the eye diagram of a receiver of the TDCM,

after conversion from the optical domain to LVDS. Fig. 12 shows the corresponding jitter and bath tub diagram. Given the relatively low speed (i.e. 400 Mbps), large eye opening and low jitter are observed.

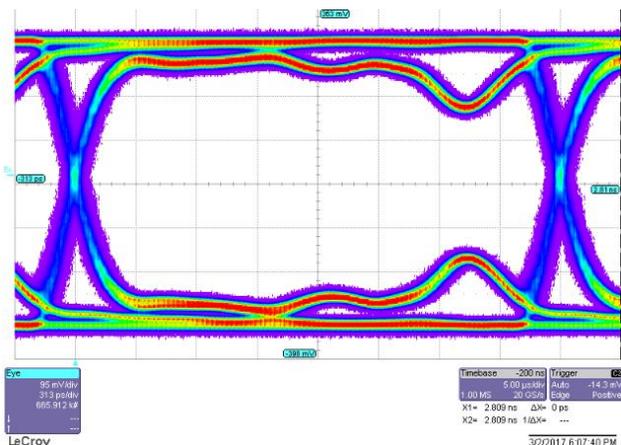

Fig. 11. Eye diagram of one of the 32 receiver ports of the TDCM (system synchronous receivers based on user I/Os). The received pattern is the link idle pattern after scrambling at 400 Mbps. The bandwidth of the oscilloscope probe is the full bandwidth (6.5 GHz).

To further assess the quality of link transmission, bit error rate testers have been embedded in the FPGA firmware of the TDCM and the ARC. This logic supports generating and verifying the correct reception of several standard pseudo-random bit patterns (PRBS7, PRBS15, PRBS23 and PRBS31) and is capable of injecting single bit errors to check that error detection works properly.

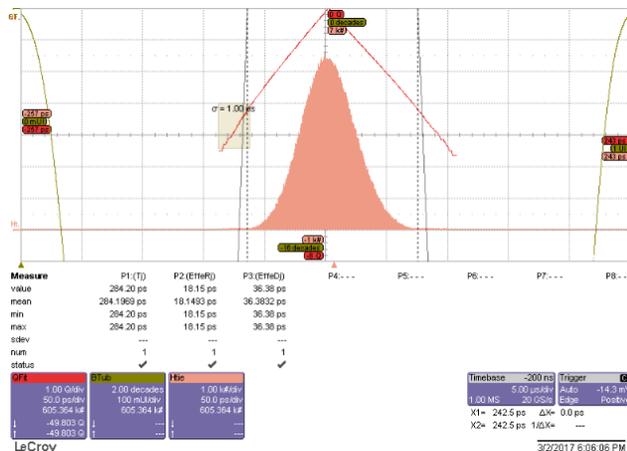

Fig. 12. Jitter and bath tub diagram of one of the 32 receiver ports of the TDCM. This corresponds to the reception of a scrambled stream at 400 Mbps.

Test have been conducted on one link between the TDCM and an ARC during 9 hours. No error has been detected in either direction. This translates into an estimated bit error rate inferior to $10^{-12}$ and $2.3*10^{-13}$ (95% confidence level) in the back-end to front-end and front-end to back-end directions respectively.

The TDCM fanout tree was tested at 500 Mbps. The receiver was a Xilinx GTP transceiver (Artix 7). This test shows the interoperability between a 100 MHz * 5 bits serializer built from a regular FPGA I/O pin and a GTP receiver (20 bit output clocked at 25 MHz). Operation at 600 Mbps (i.e. 100 MHz * 6 bit for the transmitter and 37.5 MHz * 16 bit for the GTP receiver) may be possible but requires fractional clock multiplication by 8/3 to retrieve the original 100 MHz clock. Table I summarizes the interoperability tests between the two types of transmitters and three types of receivers that are considered in this study.

TABLE I. SERIAL SPEED TESTED BETWEEN DIFFERENT TYPES OF TRANSMITTERS AND RECEIVERS. FIGURES IN PARENTHESIS ARE THE LIMITS FOR THE COMPONENTS USED: XILINX ARTIX 7 / ZYNQ SPEED 2 GRADE

|  | User I/O TX | GTP TX |
|---|---|---|
| Source synchronous User I/O RX (external ADN2815 CDR) | 100-200 Mbps (10-680 Mbps) | 50-200 Mbps (10-680 Mbps) |
| System synchronous User I/O RX (no CDR) | 400 Mbps (0-680 Mbps) | Not yet tested (0-680 Mbps) |
| GTP RX (internal CDR) | 500 Mbps (500-680 Mbps) | trivial test not done (500 Mbps-6.6 Gbps) |

Note that the range of a GTP transmitter can be extended to lower values than the minimum receive rate (500 Mbps) by sending every bit multiple times. For example, sending each bit four times at 800 Mbps leads an actual line rate of 200 Mbps, which is in the acceptable range for both a source synchronous and a system synchronous receiver based on a FPGA user I/O.

In the present era where 10 Gbps links and beyond are common, it is obviously not a remarkable technical achievement to show the operation of 100 Mbps class links, but the key aspect to note here is that these serial links are constructed using regular FPGA I/O pins.

B. *Data Collection and Event Building*

In the proposed implementation, the available bandwidth for the transport of event data from each front-end card to the TDCM is 200 Mbps and the transfer capability from the TDCM to the controlling DAQ PC is limited by the 1 Gbps bandwidth of the Gigabit Ethernet link in between.

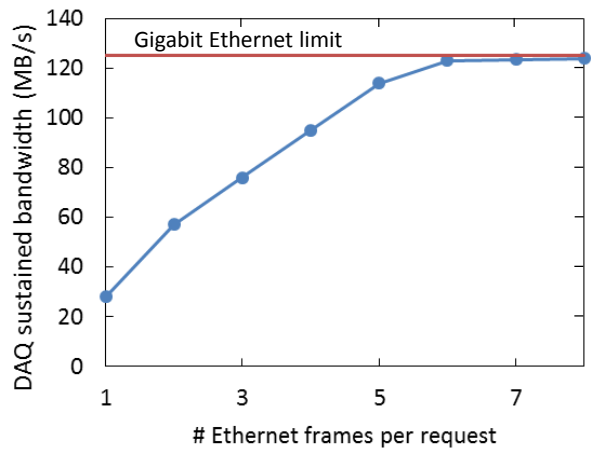

Fig. 13. Data acquisition sustained throughput for 32 emulated front-end cards varying the number of Ethernet frames that can be transmitted by the TDCM for each of the requests it receives from the DAQ PC. The Ethernet Maximum Transfer Unit (MTU) is set to 8 KB, i.e. one Jumbo frame.

Tests have been conducted with 1 and 2 ARC to acquire uncompressed events at the maximum possible speed. A data throughput of 22 MB/s and 44 MB/s is measured with one and two ARC respectively, corresponding to ~90% of the ARC-TDCM link bandwidth.

To study the operation and performance of the event builder of the TDCM before more front-end cards are available, a local event data generator is implemented in FPGA logic to emulate the presence of 1 to 32 front-end cards. Data transmission from the TDCM to the DAQ PC uses UDP/IP with a flow control mechanism where the PC can request the transfer of a given number of Ethernet frames until the next request is transmitted. The measured data acquisition throughput of the TDCM when varying the number of Ethernet frames transferable per single request is shown in Fig. 13. It can be seen that when six or more frames are allowed to be sent for each request, the maximum achievable throughput of the Gigabit Ethernet link is reached. The same tests done with the standard Ethernet MTU size (1.5 KB) show that throughput is reduced to 40 MB/s.

## VI. USE-CASES SCENARIOS

### A. Baseline Deployment

The baseline deployment of the proposed system implements the fanout structure on the TDCM carrier by making electrical signal copies of the stream to send. The TX and RX paths of the physical layer mezzanine card are used. Bi-directional multi-rate SFP transceivers allow halving the number of optical fibers compared to common dual-fiber transceivers.

### B. Reduced Material Budget Deployment

To reduce material budget on the front-end side (and possibly also background radioactivity), this deployment uses only one transmitter on the TDCM-side, e.g. a visible light LED coupled to a plastic optical fiber. On the front-end side, a discrete photo-transistor drives, after the appropriate interface, a differential pair of micro-coaxial cables connected to all front-end cards in a multi-drop LVDS chain. The return path uses a discrete LED transmitter and plastic optical fiber per front-end card. Speed is limited to ~250 Mbps per link with this type of media.

### C. Deployment for a Low Latency, Low Skew Trigger Path

Some applications may require higher bandwidth in the back-end to front-end direction, reduced latency and less dispersion on signal timing. This deployment takes one of the available high speed SERDES of the TDCM to drive the TX path of a long range SFP transceiver connected to a passive optical splitter (e.g. 1 to 8 or 1 to 16). This solution gives minimal dispersion on the timing of the distributed clock and signals. Transmission speed is not limited by the performance of FPGA regular I/O pins in this case, and 800 Mbps, 1.6 Gbps or more, can be chosen. On the front-end side, an external CDR chip is no longer adequate (or at least optimal) and a multi-gigabit SERDES embedded in the local FPGA shall be used. Published techniques can be employed to achieve deterministic latency. For transmission, the front-end side can use a regular FPGA I/O pin, or the TX path of the embedded SERDES running at an effective rate which is a sub-multiple of the serial clock rate. A PLL external to the FPGA is needed in this case to reduce the jitter of the clock recovered by the multi-gigabit SERDES so that it can be used as the reference clock of the transmitter.

## VII. CONCLUSIONS AND FUTURE WORK

This paper shows the design of a detector readout system where communications links between the front-end and back-end sides feature multiple aspects of asymmetry: topology, speed, encoding, and clocking. A fanout structure is sufficient for the back-end to front-end communication and it can be advantageous for budget material reduction. An originality of this work is that it uses regular FPGA I/O pins for serial communication at several 100 Mbps speeds. Because such I/O's are more numerous in an FPGA than dedicated high speed serial transceivers, it is possible to construct a back-end board that aggregate data from several tens of front-ends (32 in this case) in a relatively simple and cost effective way using as a commercial off-the-shelf SoC FPGA module. This architecture does not precludes that high speed SERDES are also used, and interoperability between a Xilinx GTP and user I/O-based SERDES over a certain range of speeds has been shown.

The production of the TDCM carrier and physical layer mezzanine cards will be done after conducting additional validation steps. Several units are planned to be deployed in two neutrino experiments, PandaX-III and T2K-II, along with dedicated front-end boards comparable to the ARC that are being designed by collaborators.